# Tunable Band Gap and Doping Type in Silicene by Surface Adsorption: towards Tunneling Transistors

Zeyuan Ni,[1,†] Hongxia Zhong,[1,†] Xinhe Jiang,[1,†] Ruge Quhe,[1,2] Yangyang Wang,[1] Jinbo Yang,[1,*] Junjie Shi,[1,*] and Jing Lu[1,3,*]

[1]State Key Laboratory for Mesoscopic Physics and Department of Physics, Peking University, Beijing 100871, P. R. China

[2]Academy for Advanced Interdisciplinary Studies, Peking University, Beijing 100871, P. R. China

[3]Collaborative Innovation Center of Quantum Matter, Beijing 100871, China

[†]These authors contributed equally to this work.

*Corresponding author: jinglu@pku.edu.cn, jjshi@pku.edu.cn, jbyang@pku.edu.cn

## Abstract

By using first-principles calculations, we predict that a sizable band gap can be opened without degrading its electronic properties at the Dirac point of silicene with *n*-type doping by Cu, Ag, and Au adsorption, *p*-type doping by Ir adsorption, and neutral doping by Pt adsorption. A silicene *p-i-n* tunneling field effect transistor (TFET) model is designed by adsorption of different transition metal atoms on different regions of silicene. Quantum transport simulation demonstrates that silicene TFETs have high performance featured by an on-off ratio of $10^3$, a small sub-threshold swing of 77 mV/dec, and a large on-state current over 1 mA/μm under a supply voltage of about 1.7 V. Such an on-state current is larger than that of most other TFETs. Therefore, a new avenue is opened for silicene in nano electronics.

# Introduction

Being an analogue of graphene, silicene shares both the fantastic part with graphene like the Dirac-cone-shaped energy band [1] and the ultra-high carrier mobility, theoretically of the same order of $10^5$ cm$^2$V$^{-1}$s$^{-1}$, [2] and the regretful part of the zero-band-gap nature [1]. Since the successful synthesis of silicene on Ag(111) [3-8], Ir(111) [9], and ZrB$_2$ [10], numerous efforts have been made in order to study this new member of two-dimensional (2-D) materials family. Unlike planar graphene, a band gap can be opened in low-buckled silicene by transverse electric field [11,12] or surface adsorption of alkali metal atoms [13] without degrading its electronic properties. Nonetheless, an experimentally approachable field strength can only open a gap below 0.1 eV in silicene [11,12], significantly smaller than the minimum band gap requirement (0.4 eV) for traditional field effect transistors (FETs) in a logic gate [14]. Alkali metal atom adsorption is able to induce a larger band gap up to 0.5 eV in silicene, but the alkali metal adsorbed silicene FET requires a large supply voltage $V_{dd}$ of about 30 V to turn off the device due to the high doping level.[13] Moreover, alkali metal atom adsorption can only induce *n*-type doping in silicene theoretically [13,15,16] and experimentally [17]. Modern digital logic is based on complementary metal oxide semiconductor (CMOS) technology, which requires both *n*-type and *p*-type channel metal oxide semiconductor FET (MOSFET). Power dissipation is a fundamental issue of nanoelectronic circuits. Compared with MOSFETs, tunneling FETs (TFETs) can have a smaller sub-threshold swing (SS) and supply voltage $V_{dd}$ and thus less power dissipation [18] and they requires not only *n*- and *p*-type and but also neutral doping channel (*p-i-n* junction). Additionally, optoelectronic devices such as light-emitting diode (LED) and photodiode are based on *p-i-n* junctions. Therefore, it is important to open a band gap in silicene with *p*-type and neutral doping. To the best of our knowledge, no work has been reported on the TFET based on silicene.

In this article, we explore the structural and electronic properties of silicene adsorbed by atoms of five transition metals (Cu, Ag, Au, Pt, and Ir) by using the density function theory (DFT) calculations. A sizeable band gap can be opened at the Dirac point of silicene without degrading its electronic properties after adsorption, accompanied by three doping types (*n*-type by Cu, Ag and Au adsorption, *p*-type by Ir adsorption, and neutral type by Pt adsorption). Subsequently, we propose a fabrication of *p-i-n* TFET based on silicene doped by different transition metal atoms on different regions. The transport properties of the silicene TFET is simulated by the non-equilibrium Green's function (NEGF) method coupled with the DFT or the semi-empirical (SE) Hückel approaches. The silicene TFET outperforms the proposed conventional silicene FET devices, with a very large on-state current over 1 mA/μm, an large on-off ratio of $10^3$, a small sub-threshold swing of 77 mV/dec, and a much reduced supply voltage of about 1.7 V. The on-state current even exceeds those of most of its peer TFETs.

## Computational details

The geometry optimization and electronic properties of periodic structures are performed using the projector augmented wave (PAW) method implemented in the Vienna ab-initio Simulation Package (VASP) [19-22]. The generalized gradient approximation (GGA) functional of the Perdew–Burke–Ernzerhof (PBE) form [23] and the PAW potentials [24,25] are adopted. The cut off energy is set to 450 eV after convergence tests. The supercell model consisting of a $m \times m$ unit silicene cell ($m = 1, \sqrt{3}, 2, 3,$ and 4) with one transition metal (TM) atom on the top are studied (displayed in Fig. 2) after investigating. The adsorbed silicene is denoted as TMSi$_n$, where $n = 2m \times m$. The doping concentration is defined as $N = 1/n$. An equivalent Monkhorst-Pack *k*-points grid [26] of 36 × 36 × 1 for a silicene unit cell is chosen for supercell relaxation and 40 × 40 × 1 for properties calculations. A vacuum layer of 15 Å is fixed to avoid periodic interaction. Dipole corrections perpendicular to the silicene plane is engaged in all calculations.

Both the DFT and the SE extended Hückel model coupled with the NEGF methods implemented in Atomistix Tool Kit (ATK) 11.2 are employed in the calculation of transport properties. [9,27-29] Single-ζ (SZ) and Hoffman basis are used in the DFT and the SE calculations, respectively. The same exchange-correlation functional as before is utilized in the DFT part. The *k*-points of the electrodes (channel) are set to 1 × 50 × 50 (1 × 50 × 1) and the temperature is fixed to 300 K throughout this paper. The current is calculated by using the Landauer-Büttiker formula:[30]

$$I(V_g, V_{bias}) = \frac{2e}{h} \int_{-\infty}^{+\infty} \{T_{V_g}(E, V_{bias})[f_L(E-\mu_L) - f_R(E-\mu_R)]\}dE \qquad (1)$$

where $T_{V_g}(E, V_{bias})$ is the transmission probability at a given gate voltage $V_g$ and bias voltage $V_{bias}$, $f_{L/R}$ the Fermi-Dirac distribution function for the left (L)/right (R) electrode, and $\mu_L/\mu_R$ the electrochemical potential of the L/R electrode.

## Results and discussion

### Part I: Geometry and electronic structure of silicene adsorbed by transition metal atoms

The optimized silicene structure has a Si-Si bond length of 2.279 Å and a buckling distance of 0.459 Å, which are in good agreement with the previous studies.[1,16,31,32] Four possible adsorption sites of TM atoms (Fig. 1a) are investigated: the hollow site is right above the center of a silicene hexagonal ring; the top site is above the higher Si atom; the valley site is above the lower Si atom; and the bridge site is above the middle of the Si-Si bond. The hollow site is the most preferable site of all five metals under all the checked concentrations, as illustrated in Fig. 1b. Previous calculations indicate that TM atoms like Mn, Fe, Co, Ti, and Pd also prefer hollow site.[15,16] Therefore only the hollow site configuration is studied in this work.

Structural and electronic parameters of monolayer silicene adsorbed by TM atoms are summarized in Table 1. It should be pointed out that the Dirac cones of silicene adsorbed by Au, Pt, and Ir at $N = 16.7\%$ disappear owing to the strong band hybridization. And the silicene structures are destroyed seriously when adsorbed by Pt and Ir at the highest coverage $N = 50.0\%$, as the TM atoms sink into the hexagonal ring. So we don't take these structures into consideration. When doped with Cu, Ag, and Au, the buckling height $d_0$ of silicene (Fig. 2f) ranges from 0.348 ~ 0.814 Å, generally rising with the increasing $N$. When doped with Pt and Ir, the buckling height $d_0$ of silicene ranges from 0.361 ~ 0.501 Å except for the two points at $N = 12.5\%$. At the two points, the silicene becomes almost flat, and $d_0$ is smaller than 0.06 Å. The height difference of the adsorbed TM atom and the top silicon atom $d_1$ (Fig. 2f) largely depends on the type of TM, varying from the smallest around 0.91 Å (Ir) to the largest around 1.40 Å (Ag), averaged over different $N$. On the contrary, the bond length between the TM atom and the nearest neighbor silicon atom $d_2$ is not significantly affected by the type of TM as $d_1$, averagely ranging from 2.39 Å (Ir) to 2.64 Å (Ag). Both $d_1$ and $d_2$ are insensitive to $N$ and have the same trend of Ag > Au > Pt/Cu > Ir, generally in agreement with the atomic radius order except Cu.

The adsorption energy of TM atom on silicene is defined as

$$E_a = E_{silicene} + E_T - E_c \qquad (1)$$

where $E_{silicene}$, $E_T$, and $E_{compound}$ are the relaxed energy for silicene, isolated TM atom, and the compound system, respectively. As shown in Fig. 3a, the adsorption energy is insensitive to the concentration $N$ and increases in the order of Ag < Au < Cu < Pt < Ir since smaller $d_1$ and $d_2$ generally cause a larger adsorption energy. The $E_a$ values of Cu (2.80 ~ 3.10 eV), Ag (1.80 ~ 1.96 eV), and Au (2.49 ~ 2.66 eV) are comparable to those (about 1.60 ~ 2.80 eV) of K atoms on silicene [13]. The $E_a$ values of Pt (6.14 ~ 6.24 eV), and Ir (7.66 ~ 7.75 eV) are comparable with that of W (7.05 eV [16]) on silicene. The generally large $E_a$ values of these TM atoms on silicene suggest a strong exothermic process and a lesser migration of these adsorbed TM atoms on silicene.

Fig. 4 shows the electronic structures of Cu-covered silicene at different doping levels. To distinguish the contributions between the TM atoms and Si atoms, the Si contributions are marked as different color proportional to its weight. Pure silicene has a zero gap at the Fermi level ($E_f$) (Fig. 4a). Owing to the Brillouin Zone folding, the Dirac point ($K$) of silicene is folded to the $\Gamma$ point at $n = 3x$, where $x$ is integer, including $N = 5.6\%$ (3 × 3 supercell) and 16.7% ($\sqrt{3} \times \sqrt{3}$ supercell). The Dirac cone is not destroyed by Cu adsorption but a band gap ($\Delta_D$) is opened at the Dirac point in all the examined coverages. The $\Delta_D$ generally increases with the increasing coverage, ranging from 0.03 ~ 0.66 eV (Fig. 3b), while the largest band gap in alkali-metal-covered (AM-covered) silicene is 0.50 eV [13]. The $\Delta_D$ has an extraordinary increase when the Dirac point is folded to the $\Gamma$ point. At $N = 12.5\%$ and 50.0%, there is no global band gap ($\Delta = 0$) as shown in Table 1 and Fig. 3c, and the doped silicene is actually a metal and unsuitable for a FET channel. Cu adsorption always leads to $n$-type doping of silicene (Table 1). Because the Dirac cone of silicene is not destroyed by Cu adsorption, the doping level of silicene can be qualitatively described by the Fermi level shift ($\Delta E_f$), and $\Delta E_f = E_D - E_f$, where $E_D$ is the middle energy of the band gap at the Dirac point. As given in Table 1 and Fig. 3d, the $\Delta E_f$ values are always negative. With the increasing coverage, the $|\Delta E_f|$ tends to increase (Fig. 3d), and the doping level generally increases. The work functions ($W$) of Cu-covered silicene at different coverages are summarized in Table 1 and Fig. 3e. The $W$ of pure silicene is 4.60 eV, which is in agreement with pervious study [16]. After Cu adsorption, the $W$ decreases by 0.2 ~ 0.4 eV, comparable to the $|\Delta E_f|$ values of 0.315 ~ 0.715 eV.

The influence of Cu atoms on the whole band increases with the increasing coverage from Fig. 4. The Dirac cone is dominated by Si component when $N \leq 16.7\%$ but dominated by Cu component at $N = 50.0\%$. We further provide the corresponding electron densities (Supplementary Fig. S6) of the so-called valence band maximum (VBM) and conduction band minimum (CBM) in Cu-covered silicene at the Dirac point with $N = 0$, 12.5%, and 50.0%, respectively. The electron densities of the VBM and CBM are chiefly distributed on the Si atoms at $N = 12.5\%$ while significantly extended to the Cu atoms at $N = 50.0\%$, a result consistent with the weighted band structures. We can conclude that the interaction between Cu is chiefly ionic bond at lower coverage and covalent bond component begin to appear at higher coverage.

It's well-known that the most attractive property of all graphene-like materials is the linear Dirac cone, which can lead to high mobility of charge carriers. The energy dispersion of pure silicene is linear near the Dirac point $K$, which indicates electrons and holes of silicene around the Dirac point also behave like massless Dirac fermions. We calculate the effective mass of holes ($m_h$) of Cu-covered silicene at the Dirac point through the formula:

$$m_h = (\frac{1}{\hbar^2}\frac{d^2E}{dk^2})^{-1} \qquad (2)$$

The $m_h$ values are anisotropic under external electric field [11] but are isotropic after Cu adsorption. The $m_h$ averaged over the different directions as a function of coverage is given by Fig. 3f. Of all the examined Cu-covered silicene, the $m_h$ is small ranging from 0.01 ~ 0.16 $m_0$ ($m_0$ is the free electron mass). The $m_h$ and $\Delta_D$ share the same change tendency with the coverage and peak at $n = 3x$, proving that there is a trade-off relationship between $m_h$ and $\Delta_D$. Among the five coverages in Cu-covered silicene, $N = 5.6\%$ and $16.7\%$ are the best choice for TFET because they realize a relatively large direct overall band gap of $\Delta = 0.18$ and $0.66$ eV while preserving a small effective mass of $0.07\ m_0$ and $0.16\ m_0$ simultaneously.

The electronic band structures of silicene adsorbed by the five kinds of TM atoms at $N = 5.6\%$ are displayed in Fig. 5. They share similar band dispersions and direct band gaps ($0.14 \sim 0.23$ eV) at the Dirac point, but with different doping levels: Ag and Au adsorption lead to $n$-type doping of silicene, Pt adsorption leads to neutral doping, and Ir adsorption leads to $p$-type doping as also summarized in Table 1. We notice that N adsorption, B and Al substitution also cause a band gap opening ($\sim 0.1$ eV) at the Dirac point with $p$-type doping in silicene without destroying the Dirac cone in a recent DFT calculation.[31] The Dirac point is mainly contributed by Si atoms when adsorbed by Ag and Pt, like Cu, but a small amount of TM component is available in the case of Au and Ir adsorption.

The electronic structures of other doped silicene are provided in Fig. S1-4. The Dirac cones of Pt and Ir-covered silicene with $N = 16.7\%$ have been destroyed and unsuitable for high performance FET. The $\Delta_D$ values of Ag, Au, Pt, and Ir-adsorbed silicene at different coverages are presented in Fig. 3b. The change tendencies of the band gap of Ag, Au, Pt, and Ir-covered silicene as a function of coverage and the band gap size at the same coverage are similar to those of Cu-covered silicene, and the band gap values range from $0.006 \sim 0.49$ eV. At $N = 12.5\%$ and $50.0\%$, there is no band gap in the whole energy range for Ag and Au-covered silicene because the TM-derived band overlaps the Dirac cone in energy. The maximum global band gap of $p$-doped and neutral doped silicene without degrading its electronic properties are $0.222$ (IrSi$_{18}$) and $0.23$ eV (PtSi$_{18}$) with hole effective mass of $0.13\ m_0$ and $0.085\ m_0$, respectively.

The $\Delta E_f$ and $W$ of Ag, Au, Pt, and Ir-adsorbed silicene at different coverages are presented in Table 1 and Fig. 3d-e. The $\Delta E_f$ values of Au and Ag-adsorbed silicene are always negative, the $\Delta E_f$ values of Pt-adsorbed silicene are zero, and the $\Delta E_f$ values of Ir-adsorbed silicene are always positive, demonstrating the unchanged doping type of TM-covered silicene with the covergae. With the increasing coverage, the $|\Delta E_f|$ values of Ag, Au, Pt, and Ir-adsorbed silicene tend to increase slowly. Given the same coverage, the $|\Delta E_f|$ values of $n$-type doped

silicene decrease in this order: Cu > Ag > Au. Moreover, the $|\Delta E_f|$ values of Au-covered silicene are about 0.2 eV, smaller than those of Ir-covered silicene in all examined coverages. When $\Delta E_f < 0$ (*n*-type doping), the corresponding $W$ always decreases; When $\Delta E_f > 0$ (*p*-type doping), the corresponding $W$ always increases. Although $\Delta E_f = 0$ (neutral doping), the $W$ values of Pt-covered silicene are slightly larger than that of pure silicene. The differences of $W$ between TM-adsorbed (TM = Ag, Au, Pt, and Ir) and pure silicene are comparable to the corresponding $|\Delta E_f|$ values in all examined coverages as the case of Cu-covered silicene.

The values of $m_h$ for Ag, Au, Pt, and Ir-adsorbed silicene at different coverages are presented in Fig. 3f. The variations of $m_h$ of TM-adsorbed silicene (TM = Ag, Au, Pt, and Ir) with the coverage are similar to that of Cu-adsorbed silicene, and the values range from 0.01 ~ 0.16 $m_0$. Based on the fact that bilayer graphene has an extremely high carrier mobility of $2\times10^5$ cm$^2$V$^{-1}$s$^{-1}$ on suspended sample with effective mass of $m_e = 0.03\ m_0$ [33], the hole mobilities in the checked TMSi$_n$ with an undestroyed Dirac cone are estimated to be $4 \times 10^4$ ~ $6 \times 10^5$ cm$^2$V$^{-1}$s$^{-1}$, assuming that the scattering time of TMSi$_n$ is the same as suspended graphene.

Spin-orbit coupling (SOC) can affect the band structure of silicene. A direct band gap of 1.48 meV appears at the Dirac point of pure silicene due to SOC effects, which is in excellent agreement with previous works [13,16,31]. We compare the band structures with and without the inclusion of the SOC effects of silicene adsorbed by the five kinds of TM atoms at $N = 5.6\%$ in Fig. S5. The band dispersions of TMSi$_{18}$ (TM = Cu, Ag, Au, Pt, and Ir) with and without the inclusion of the SOC effects are nearly the same, but the opened gap at the Dirac point decrease by 0.7, 7, 30, 30, and 44 meV, respectively. Therefore, the electronic band structures we present are believable.

Finally, we discuss the mechanism of band gap opening in silicene after adsorption. When silicene is adsorbed by Cu, Ag, Au, and Ir ($\Delta E_f \neq 0$), there exists charge transfer between TM atoms and Si atoms, producing a built-in electric field (as shown in Fig. S7), which breaks the inversion symmetry in silicene and induces a band gap. Even if $\Delta E_f = 0$ (i.e. no charge transfer), there also exists a band gap in Pt-adsorbed silicene. Previous studies have indicated that even if the doping level is zero, a band gap of 0.05 and 0.093 eV is induced in bilayer graphene and ABC-stacked trilayer graphene, respectively, adsorbed on metal substrates. [34]

The reason is assigned to the inversion symmetry breaking due to the wave function interaction between metal and the bottom graphene [34]. The bucked structure of monolayer silicene is similar to the structure of bilayer graphene. Therefore, the wave function interaction between Pt and the bottom Si layer breaks the inversion symmetry of silicene and opens the band gap in Pt-adsorbed silicene. However, the two physical models could not explain the fact that TMSi$_{3x}$ have notably larger band gaps. Former study has suggested that the breaking of bond symmetry is responsible for the exceptionally large band gap in AMSi$_{3x}$ [13]. We attribute the exceptionally large band gap in TMSi$_{3x}$ to the same the mechanism of bond symmetry broken.

## Part II: Silicene tunneling field effect transistor

The entire *p-i-n* silicene TFET model is displayed in Fig. 6. The source, drain, and channel are doped with Ir, Cu, and Pt to open a band gap with *p*, neutral, and *n*-type doping, respectively. The structures of the three isolated regions are taken from the optimized ones obtained by VASP. As mentioned above, the adsorption energies $E_a$ of Ir, Cu, and Pt are relatively larger than alkali metals by 1 ~ 4 eV and thus are more stable. The doping concentration *N* of all the regions is fixed at 5.6%, resulting in similar band gaps of about 0.2 ~ 0.3 eV in all the regions in both the DFT and the SE calculations (See Supplementary Materials). There is a single gate lying over the channel with a dielectric region made of HfO$_2$ (dielectric constant $k$ = 25 [35]). A few layers of hexagonal boron nitride (*h*-BN) are placed under the dielectric to protect the silicene from being destroyed by the oxide [11]. The transport properties are mainly calculated by the SE approach, which are benchmarked with the DFT approach in Fig. 7 and in Supplementary Materials.

First, a TFET with a channel length $L$ = 2 nm under zero gate voltage is investigated. Both the DFT and the SE results show an apparent NDR effect in its $I$-$V_b$ curve displayed in Fig. 7 (a)), typical of Esaki diode [36]. The peak-to-valley ratios (PVRs) calculated by the DFT (PVR = 1.6) and the SE (PVR = 1.8) methods are very close. Transmission coefficient is proportional to the product of density of states (DOS) of electrodes and channel: [37,38]

$$T(E) \propto D_S(E)D_C(E)D_D(E) \qquad (3)$$

where $D_S$, $D_C$, $D_D$ stands for the DOS of the three parts of a FET device: source, channel and drain, respectively. Thus, a gap in the DOS of any region will lead to a gap in the transmission spectrum of the same size and position. With the fact in mind, let's focus on the DFT transmission spectra of $L = 2$ nm TFET under different bias voltages (Fig. 8). There are two obvious gaps around $E_f$ in the transmission spectrum in $V_b = 0$ case (Fig. 8(a)). The gap above $E_f$ (blue arrow) is induced by the Ir adsorption in the source electrode, since its width (~ 0.3 eV) and position (at 0.4 eV) are close to those of the band gap in the Ir-adsorbed silicene of the same concentration shown in Fig. 5 and Supplementary Materials. Likewise, the gap below $E_f$ (at -0.6 eV with a width of ~ 0.2 eV) is caused by the Cu adsorption in the drain electrode. As $V_b$ increases, not only is the bias window expanded, but the distance of the two gaps is also made close due to the change of the chemical potential of source and drain. The total current is a trade-off between the expansion of the bias window, which favors the current, and the increasing magnitude of the gap within the bias window, which suppresses the current. From Eq. 3, we know that the total current is dominated by the integration of the transmission coefficient times the Fermi distributions of the electrodes over all energy range. Thanks to the Fermi distribution, the major contribution to the total current is made by transmission spectrum within the bias window. The transmission outside the bias window does have minor effect on the total current due to the "tail" of the Fermi distribution, but is not important in this case. As a result, the total current reaches a peak value when the Ir-related gap just begin to enter the bias window ($V_b = 0.3$ V, Fig. 8(b)) and then drops to a valley when this gap is almost included in the bias window ($V_b = 0.5$ V, Fig. 8(c)). Continually increasing the bias (e.g. in Fig. 8(d)), the valence band of the source enters the bias window and the current restores to increase.

If the channel length is increased to $L = 4$ nm (Fig. 7 (b)), the PVR drops to ~1.1. Such a reduced PVR with the increasing $L$ in silicene TFET is in accordance with previous researches on the graphene and carbon nanotube $p$-$n$ junction [39,40]. In fact, such a degeneration is caused by the widening of the central barrier with the increasing channel length. Fig. 9 shows the transmission spectra of the TFETs with different $L$ under $V_b = 0.1$ V and zero $V_g$. When $L$ is as short as 2 nm, the gap in the channel is too thin to be reflected in the spectrum. As a result, the transmission coefficients around $E_f$ are very high, leading to a large current.

With the channel lengthened, the transmission coefficients around $E_f$ are gradually suppressed by the appearing channel gap induced by Pt doping, thus causing a decreasing total current. Note that such length effect affects majorly on the peak current. In the case of the valley current, the source gap is located around the position of the channel gap (Fig. 8c and Fig. 9), there is only one visible gap in the transmission spectra whatever $L$ is, and the valley current remains almost unaffected. Consequently, NDR effect fades as the peak current goes down, and the valley current stays the same when the channel is lengthened.

The appearing channel gap provides the possibility of current switching in our TFET. Next we use the SE approach to study the long channel TFETs, which are beyond the capability of the DFT approach. Although there is a discrepancy in the peak/valley positions between the DFT and the SE results in Fig. 7, which is caused by the difference of $E_f$ of doped silicene calculated by the two methods (See Supplementary Materials), the corresponding current densities are of the same order of magnitude. Besides, the PVRs calculated by the DFT approach are almost the same as those by the SE approach in both $L = 2$ nm and 4 nm TFETs. Thus, the SE approach appears to be a good substitution of the DFT approach in the transportation calculation of our silicene TFET model at this coverage.

While the gate voltage has little effect on the TFETs with $L$ shorter than 4 nm, the $I_{ds}$-$V_g$ characteristics of those with larger $L$ display an apparent switching effect, which is typically shown in Fig. 10(a) for the $L = 16$ nm TFET. The curve minimum, or the off state, is located at $V_g = 0.27$ V. If $V_g = 2$ V is chosen as the on-state, the on-off ratio can reach over $10^3$ within a supply voltage $V_{dd} = 1.73$ V, and the steepest SS is 90 mV/dec. The transmission eigenstates at $E_f$ and the $k$-point $\Gamma(0,0)$ of the off- and on-state are given in Fig. 10(b) and (c), respectively. Apparently the transmission eigenstate cannot reach the drain from source when the device is turned off (Fig. 10(b)), while the transmission eigenstate is able to connect the two leads when the device is turned on (Fig. 10(c)). Notice that the on-state current is over 1.8 mA/μm, larger than most of the common TFETs [18,41-47]. Such an on-state current also meets the requirement of $I_{on} = 1.6$ mA/μm for 2016 high performance FETs with $L = 15.3$ nm proposed in the 2012 International Technology Roadmap for Semiconductors (ITRS) [48]. It is well known that a large on-state current is beneficial to shorten the gate delay and speed up the device operation [49].

To illustrate the source of the switching effect, the transmission spectra of the $L = 16$ nm TFET under different $V_g$ are provided in Fig. 11. When there is no gate voltage applied, the channel gap induced by Cu adsorption is just above $E_f$ due to slight doping effect by the electrodes, as is shown in Fig. 11(a). Fig. 11(b) shows the transmission spectrum under $V_g = 0.27$ V. The center of the channel gap (green arrow) is moved to $E_f$, making the entire bias window "blocked", and the transmission coefficients are very low within it. As a result, the corresponding current reaches a minimum, and the device is turned off. If we keep on adding up $V_g$, the channel gap will be gradually pushed away from the bias window, leading to an increasing current. When the gap is completely moved out, like the case at $V_g = 2$ V in Fig. 11(c), the transmission coefficients within the bias window grow dramatically, leading to a large on-state current. Note that the gaps from the source and drain are static, since $V_g$ is only applied to the channel.

The relation of the on-off ratio and the steepest SS to the channel length $L$ is displayed in Fig. 12. The on-off ratio rises monotonically with $L$ and exceeds $10^3$ when $L \geqslant 16$ nm, coordinate with the evolution of the channel gap size drawn in Fig. 9. This is easy to understand, since larger $L$ leads to deeper channel gap and lower off-state current, thus resulting in a better on-off ratio. The steepest SS generally has the opposite trend but has a minimum of 77 mV/dec at $L = 32$ nm and becomes slightly larger of about 86 mV/dec at $L = 48$ nm. Both the on-off ratio and SS degenerate dramatically when $L$ enters sub-10-nm region and saturates when $L$ is beyond 32 nm, indicating that the best performance could be achieved around 10 – 30 nm. Greater on-off ratio of $10^4$, larger on-state current over 2 mA/μm, and smaller $V_{dd}$ below 1 V may be realized if the device is further optimized.

Alternatively, a *p-i-n* silicene TFET could be made if one uses three pairs of dual-gate on both the leads and the channel of a pristine silicene TFET, which allow it to open a band gap in silicene with different doping levels in different regions. Such a silicene TFET is not only more complicate due to the requirement of six gates but also difficult to reach a high current on-off ratio because the maximum band gap under an experimentally accessible electric field is less than 0.1 eV [11,12].

# Conclusion

In summary, the geometric and electronic properties of silicene adsorbed by transition metal atoms (Cu, Ag, Au, Pt, and Ir) are investigated by first-principles calculations. All dopants favor the same configuration on silicene and induce a sizable band gap at the Dirac point without degrading the electronic properties at lower coverage. Importantly, three doping types are observed for the first time for silicene, which enables the simulation of a silicene *p-i-n* TFET by three types of doping in different regions of silicene. We predict that the silicene TFET outperforms traditional silicene FETs by an on-state current over 1 mA/μm, which is even larger than most of its peer TFETs, and an on-off ratio of $10^3$, a subthreshold swing of 77 mV/dec, and a small supply voltage of about 1.7 V. The on-state current is also of the same order of magnitude as the latest requirement of ITRS on 2016 high performance FET. In addition, the ultrathin silicene *p-i-n* junction with a direct band gap also sheds light on high-efficiency optoelectronic devices. Hence, new prospects are opened up for silicene in nano electronics and optoelectronics.

# Acknowledgements

This work was supported by the National Natural Science Foundation of China (Nos. 11274016, 51072007, 91021017, 11047018, and 60890193), the National Basic Research Program of China (Nos. 2013CB932604 and 2012CB619304), Fundamental Research Funds for the Central Universities, National Foundation for Fostering Talents of Basic Science (No. J1030310/No. J1103205), Program for New Century Excellent Talents in University of MOE of China, in part by the Materials Simulation Center, a Penn-State MRSEC and MRI facility.

# Author contributions

The idea was conceived by J. L. The DFT electronic band calculation (VASP part) was performed by H. Z., X. J., and R. Q., and the SE electronic band calculation and the device simulation was performed by Z., N. and Y., W. The data analyses were performed by J. L., J. Y., J. S., Z. N., H. Z., R. Q., Y. W. This manuscript was written by Z. N., H. Z., J. Y., and J. L.

All authors review this manuscript.

# Additional information

**Supplementary Materials** accompanies this paper at http://www.nature.com/scientificreports

**Competing financial interests:** The authors declare no competing financial interests.

Table 1: Calculated parameters for Cu, Ag, Au, Pt and Ir-adsorbed silicene at different coverage $N$: the silicene buckling ($d_0$), the height between adsorption atom and top-surface silicene ($d_1$), the bond length between adsorption atom and top-surface silicene ($d_2$), band gap at the Dirac point ($\Delta_D$), global band gap ($\Delta$), Fermi level shift ($\Delta E_f$), work function of the stable structure ($W$), doping type: negative ($n$), intrinsic ($i$), and positive ($p$), respectively. The calculated $d_0$ and $W$ of pure silicene are 0.459 Å, and 4.6 eV, respectively. At $N = 16.7\%$, the Dirac cones of silicene adsorbed by Au, Pt, and Ir are destroyed. At $N = 50.0\%$, the silicene structures are destroyed seriously when adsorbed by Pt and Ir as the TM atoms sink into the hexagonal ring. We therefore don't take these structures into consideration.

|    | Coverage (%) | $d_0$ (Å) | $d_1$ (Å) | $d_2$ (Å) | $\Delta_D$ (eV) | $\Delta$ (eV) | $\Delta E_f$ (eV) | $W$ (eV) | $p/i/n$ |
|----|------|-------|-------|-------|-------|-------|--------|-------|---|
|    | 3.1  | 0.349 | 0.958 | 2.434 | 0.035 | 0.035 | -0.360 | 4.214 | $n$ |
|    | 5.6  | 0.382 | 0.953 | 2.423 | 0.180 | 0.180 | -0.398 | 4.211 | $n$ |
| Cu | 12.5 | 0.554 | 1.154 | 2.486 | 0.060 | 0.000 | -0.315 | 4.382 | $n$ |
|    | 16.7 | 0.578 | 0.937 | 2.431 | 0.660 | 0.660 | -0.515 | 4.220 | $n$ |
|    | 50.0 | 0.814 | 1.021 | 2.454 | 0.190 | 0.000 | -0.715 | 4.315 | $n$ |
|    | 3.1  | 0.348 | 1.362 | 2.629 | 0.030 | 0.030 | -0.368 | 4.225 | $n$ |
|    | 5.6  | 0.401 | 1.377 | 2.628 | 0.140 | 0.140 | -0.360 | 4.203 | $n$ |
| Ag | 12.5 | 0.446 | 1.402 | 2.638 | 0.020 | 0.000 | -0.322 | 4.287 | $n$ |
|    | 16.7 | 0.571 | 1.384 | 2.636 | 0.490 | 0.490 | -0.464 | 4.197 | $n$ |
|    | 50.0 | 0.723 | 1.446 | 2.659 | 0.010 | 0.000 | -0.368 | 4.292 | $n$ |
|    | 3.1  | 0.388 | 1.232 | 2.553 | 0.030 | 0.030 | -0.234 | 4.329 | $n$ |
| Au | 5.6  | 0.477 | 1.279 | 2.557 | 0.180 | 0.180 | -0.201 | 4.403 | $n$ |
|    | 12.5 | 0.502 | 1.283 | 2.554 | 0.030 | 0.000 | -0.219 | 4.372 | $n$ |
|    | 50.0 | 0.710 | 1.283 | 2.575 | 0.250 | 0.000 | -0.196 | 4.547 | $n$ |
|    | 3.1  | 0.361 | 0.984 | 2.424 | 0.030 | 0.030 | 0.000  | 4.618 | $i$ |
| Pt | 5.6  | 0.382 | 0.982 | 2.418 | 0.230 | 0.230 | 0.000  | 4.763 | $i$ |
|    | 12.5 | 0.037 | 1.071 | 2.512 | 0.020 | 0.020 | 0.000  | 4.725 | $i$ |
|    | 3.1  | 0.501 | 0.893 | 2.355 | 0.006 | 0.006 | 0.365  | 4.948 | $p$ |
| Ir | 5.6  | 0.486 | 0.940 | 2.365 | 0.222 | 0.222 | 0.362  | 5.056 | $p$ |
|    | 12.5 | 0.056 | 0.900 | 2.444 | 0.040 | 0.040 | 0.569  | 5.137 | $p$ |

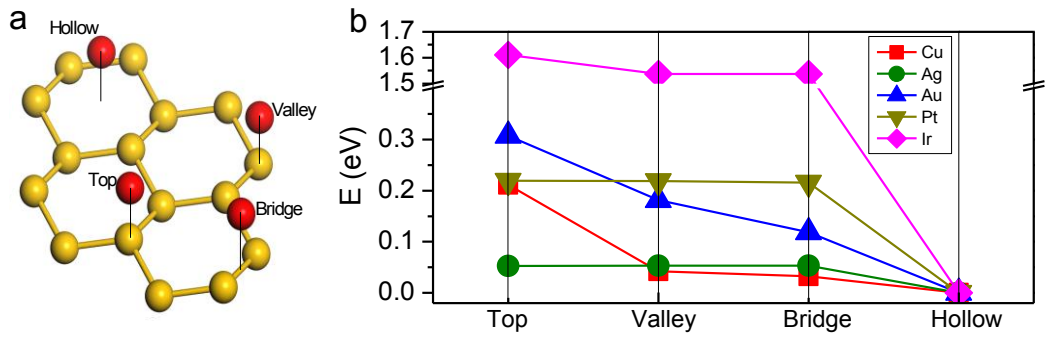

Fig. 1: (a) Four possible adsorption sites considered. (b) Total energy relative to the hollow site for the four adsorption sites.

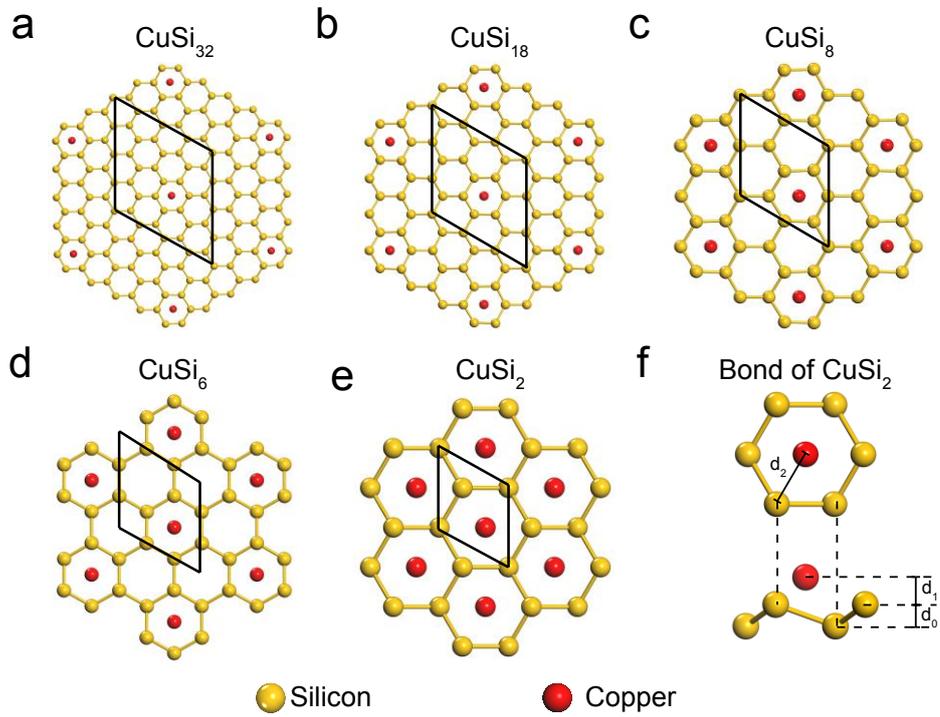

Fig. 2: Structures of Cu-covered silicene with coverage $N$ = 3.1% (a), 5.6% (b), 12.5% (c), 16.7% (d), and 50.0% (e), respectively. The rhombi plotted in black line shows the unit cell for each structure. (f) Top and side view of Cu-covered silicene supercell. $d_0$: the silicene buckling; $d_1$: the height between the adsorption atom and top-surface silicene; $d_2$: the bond length between the adsorption atom and top-surface silicene.

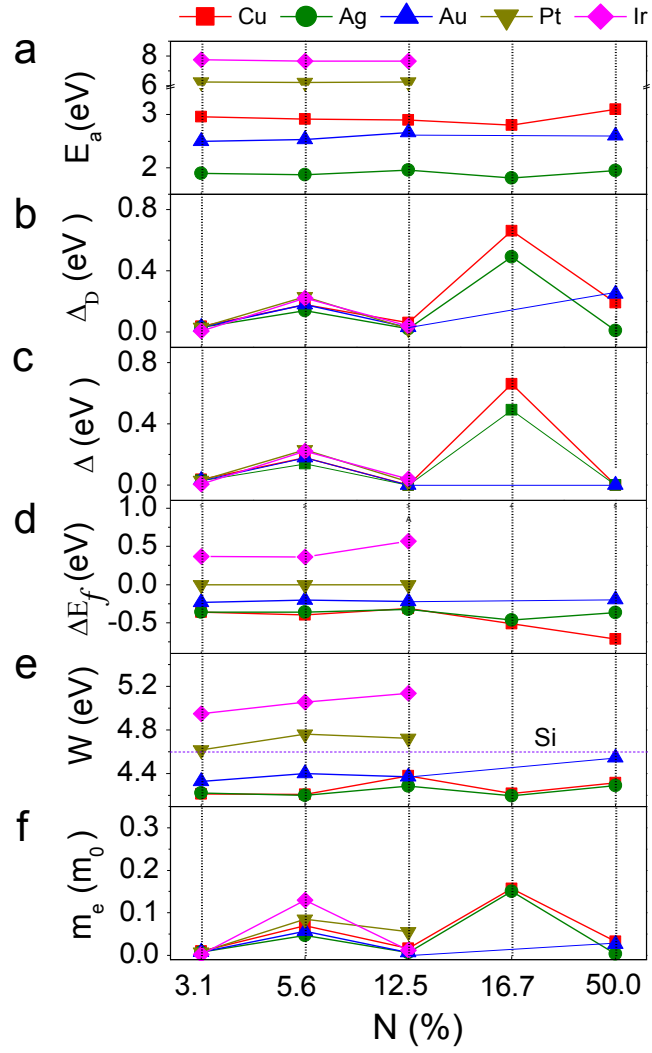

Fig. 3: (a) Adsorption energy (per metal atom), (b) band gap at the Dirac point, (c) global band gap, (d) Fermi level shift of metal covered-silicene, (e) Work function, the horizontal dashed line stands for the work function of pure silicene, and (f) effective mass of holes of the metal covered silicene as a function of coverage.

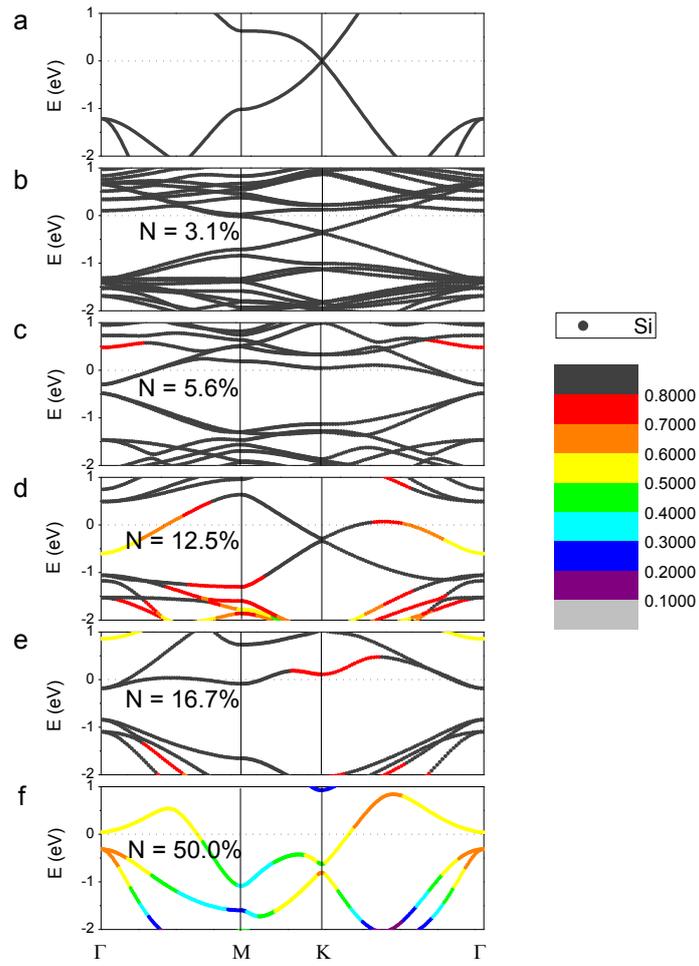

Fig. 4: (a-f) Electronic band structures of the Cu-covered silicene at coverages of $N = 0$, 3.1%, 5.6%, 12.5%, 16.7%, and 50.0%, respectively. The Fermi level is set to zero. Contributions from the silicon atoms are marked as different color proportional to the weight.

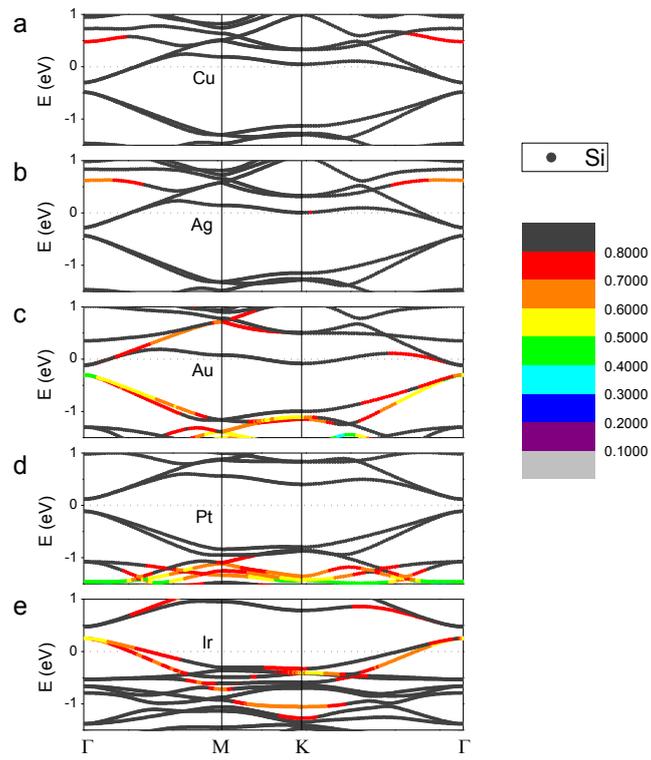

Fig. 5: Electronic band structures of (a) Cu, (b) Ag, (c) Au, (d) Pt, and (e) Ir-covered silicene at the coverage of $N = 5.6\%$. The Fermi level is set to zero. Contributions from the silicon atoms are marked as different color proportional to the weight.

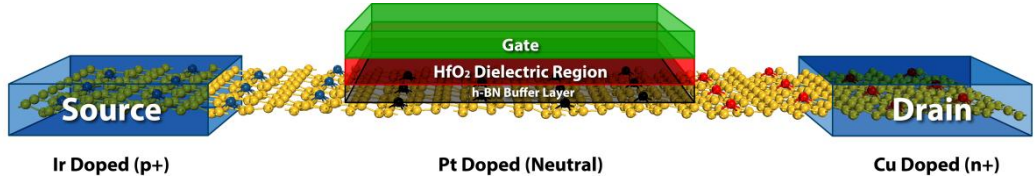

Fig. 6: Schematic model of the silicene TFET. The yellow silicon atoms form a sheet of monolayer silicene. Source/Drain is Ir (blue atoms)/Cu (red atoms) doped silicene and thus is $p+/n+$ type semiconductor with a band gap of about 0.3/0.2 eV. The central region, or the channel, is doped with Pt (black atoms) and thus is a neutral type semiconductor with a gap around 0.3 eV. The concentration of the dopants in all parts of the device is fixed at $N = 5.6\%$ (TMSi$_{18}$, TM = Cu, Pt, and Ir). A 0.5 nm thick HfO$_2$ dielectric region ($k = 25$) is placed over the channel. A thin h-BN buffer layer is utilized to protect the silicene from the oxide.

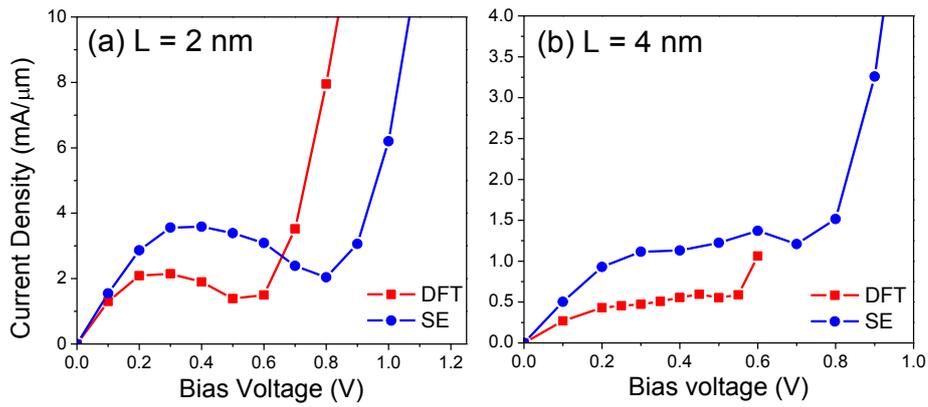

Fig. 7: Calculated current density versus bias voltage of the silicene TFETs with different channel lengths: (a) $L = 2$ nm and (b) $L = 4$ nm. Red and blue lines denote the DFT and the SE data, respectively. The PVR by the DFT/SE approach is ~ 1.6/1.8 for $L = 2$ nm device and 1.1/1.1 for $L = 4$ nm device.

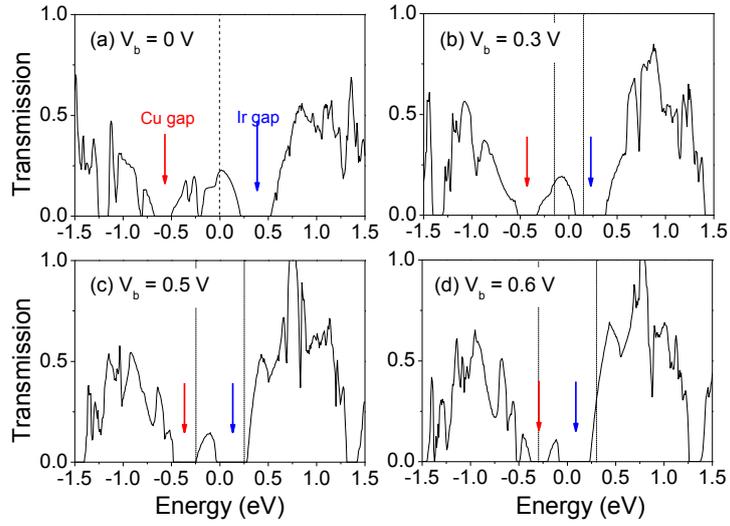

Fig. 8: DFT transmission spectra of the $L = 2$ nm silicene TFET under different $V_b$: (a) $V_b = 0$ V (b) $V_b = 0.3$ V (at peak) (c) $V_b = 0.5$ V (at valley) (d) $V_b = 0.6$ V (after valley). Red and blue arrows indicate the gaps induced by Cu in drain and Ir in source, respectively. The dashed lines indicate the bias window. As the $V_b$ increases, the bias window is widened and the two gaps approach each other. When the Ir gap is moved into the bias window, NDR happens.

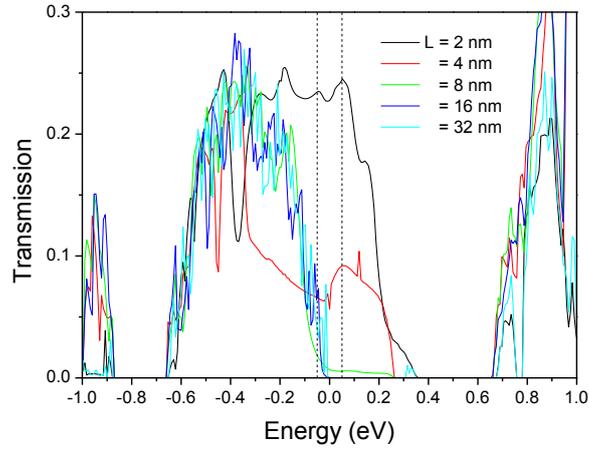

Fig. 9: SE transmission spectrum of the silicene TFETs with different channel lengths under $V_g = 0$ V and $V_b = 0.1$ V. The dashed lines indicate the bias window. Same as in Fig. 8, the untouched gaps at -0.8 eV and 0.5 eV are drain and source gaps, respectively. Note that as the channel length increases, the channel gap induced by Pt gradually appears around 0.1 eV and lowers the transmission coefficients within the bias window. As a result, the peak current is suppressed, and NDR effect fades.

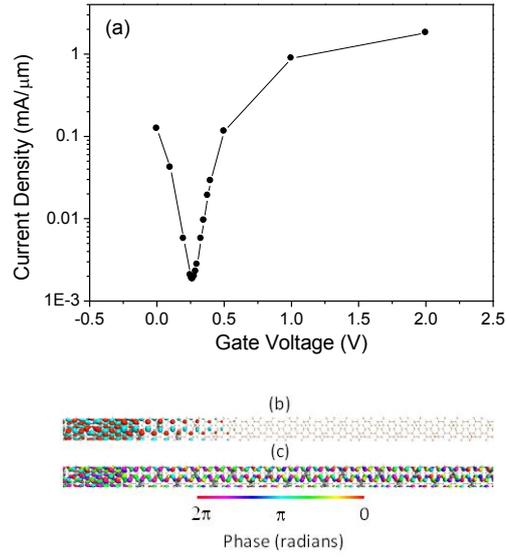

Fig. 10: (a) Current density versus gate voltage of the $L = 16$ nm silicene TFET at $V_b = 0.1$ V by the SE approach. The on-off ratio is about $10^3$, and the steepest SS is about 90 mV/dec. (b) and (c): the transmission eigenstate of the same model at $E_f$ and the $k$-point $\Gamma(0, 0)$ under different $V_g$: (b) $V_g = V_{off} = 0.27$ V and (c) $V_g = V_{on} = 2.0$ V. The isovalues of the isosurfaces in (b) and (c) are fixed to 0.1 au.

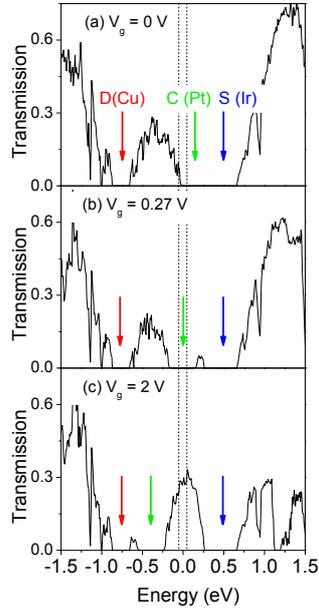

Fig. 11: SE transmission spectra of the $L$ = 16 nm silicene TFET at $V_b$ = 0.1 V under different $V_g$: (a) $V_g$ = 0, (b) $V_g$ = $V_{off}$ = 0.27 V, and (c) $V_g$ = $V_{on}$ = 2.0 V. The transmission gaps induced by different metal atoms in the separate parts of TFET are indicated by arrows of different colors, with D/C/S standing for the drain/channel/source region of the device, respectively. As is displayed, the gaps of the leads (i.e. D and S) are unchanged, while the channel gap moves as $V_g$ changes. The device is turned off when the channel gap crosses the entire bias window at $V_g$ = 0.27 V and is turned on when the channel gap is moved away at $V_g$ = 2 V.

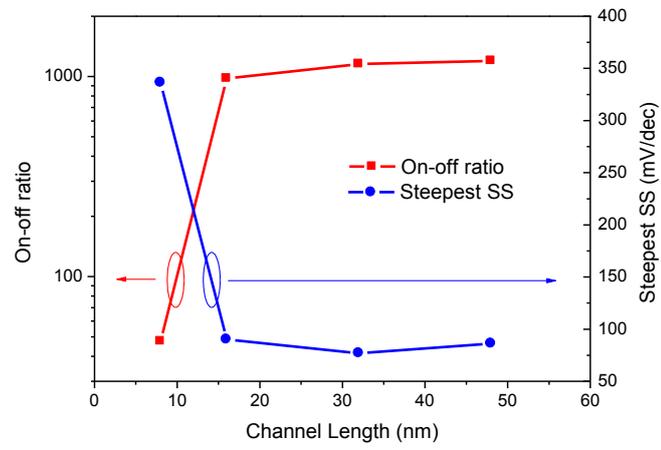

Fig. 12: Current on-off ratio (red) and steepest SS (blue) of the silicene TFETs with different channel lengths by the SE approach.